\documentstyle[11pt,newpasp,twoside,epsf]{article}
\begin{document}
\title{Strong Limits on the Infrared Spectrum of HD\,209458\,b Near 2.2 \micron}

\author{L. Jeremy Richardson \emph{and} Drake Deming}
\affil{NASA's Goddard Space Flight Center, Code 693, Greenbelt, MD  20771}
\author{Sara Seager}
\affil{Department of Terrestrial Magnetism, Carnegie Institution of Washington, 5241 Broad Branch Road, NW, Washington, DC  20015}

\begin{abstract}
We present a brief summary of observations of the transiting
extrasolar planet, HD\,209458\,b,
designed to detect the secondary eclipse.  We employ the method of
`occultation spectroscopy', which searches in combined light (star and planet)
for the disappearance and
reappearance of weak infrared spectral features due to the planet as it passes
behind the star and reappears. 
We have searched for a continuum peak
near 2.2 \micron\ (defined by CO and H$_2$O absorption bands),
as predicted by some models of the planetary
atmosphere to be $\sim 6 \times 10^{-4}$ of the stellar flux, 
but no such peak is detected at a level of $\sim 3 \times
10^{-4}$ of the stellar flux.
Our results represent the strongest limits on the infrared spectrum
of the planet to date and carry significant implications for
understanding the planetary atmosphere. 
\end{abstract}

\section{Background}
Our observations cover two predicted
secondary eclipse events, and we
obtained 1036 individual spectra of the HD\,209458 system using the SpeX
instrument at the NASA IRTF in September 2001.  Our spectra 
extend from 1.9 to 4.2 \micron\ with a
resolution ($\lambda/\Delta \lambda$) of 1500.  A summary of the
method of occultation spectrosocpy, as well as the details of the
data analysis, can be found in Richardson, Deming, \& Seager (2003).

\section{Results and Discussion}
The final difference spectrum, shown in Figure~1,
represents the average candidate planetary spectrum, as calculated from the
`in-eclipse' minus the `out-of-eclipse' spectra from the two nights
during which a secondary eclipse was predicted to occur.  The average
spectrum represents data from 550 individual spectra of HD\,209458, as
well as an equal number of spectra of the comparison star HD\,210483.
The comparison star was used in our analysis to remove variability due
to changes in the terrestrial atmsophere and to normalize the data to
the stellar flux density.  Also shown in Figure~1 is the baseline
model for HD\,204958\,b calculated by Sudarsky, Burrows, \& Hubeny
(2003), which exhibits a peak near 2.2 \micron.   A
least-squares analysis indicates this peak is not present in the
candidate planetary spectrum as derived from the data.

We believe this result has significant implications for the structure
of the planetary atmosphere.
In particular, some models that assume the stellar irradiation is
re-radiated entirely on the sub-stellar hemisphere predict this flux
peak, which is
inconsistent with our observations.
Several physical mechanisms can
improve agreement with our observations, including the re-distribution of
heat by global circulation, a nearly isothermal atmosphere, and/or the
presence of a high cloud.

\begin{figure}[!t]
\epsfysize=3.7in
\plotone{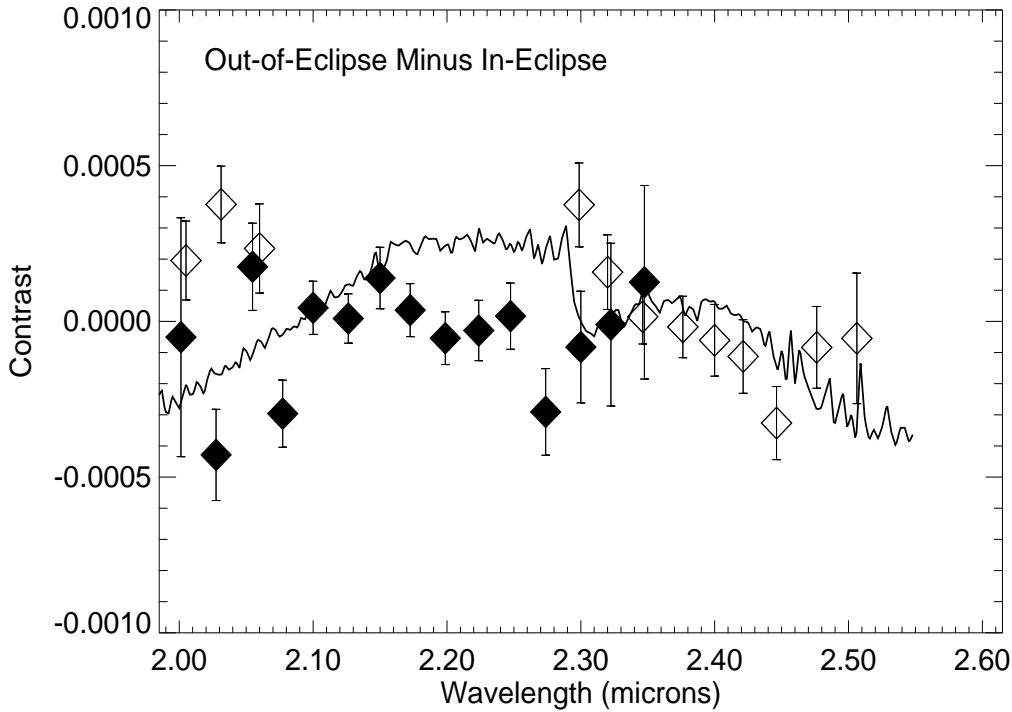}
\caption{Final candidate planetary spectrum (symbols).  Solid line represents
  the baseline model for HD\,209458\,b as caluclated by Sudarsky,
  Burrows, \& Hubeny (2003).
The offset between the model
and the data is not important, since we are comparing the shapes of
the two spectra; thus, the mean has been subtracted from the
data and model, respectively, for plotting purposes.
\label{fig:sw}}
\end{figure}

\end{document}